# Metacomposite characteristics and their influential factors of polymer composites containing orthogonal ferromagnetic microwire arrays

Y. Luo[1], H.X. Peng[1,a)], F.X. Qin[2,b)], M. Ipatov[3], V. Zhukova[3], A. Zhukov[3], J. Gonzalez[3]

[1]Advanced Composite Centre for Innovation and Science, Department of Aerospace Engineering, University of Bristol, University Walk, Bristol, BS8 1TR, UK

[2]1D Nanomaterials Group, National Institute for Material Science, 1-2-1 Sengen, Tsukuba, Ibaraki 305-0047, Japan

[3]Dpto. de Fisica de Materiales, Fac. Quimicas, Universidad del Pais Vasco, San Sebastian, 20009, Spain

The microwave properties of glass-fibers reinforced polymer composite embedded with an orthogonal array of $Fe_{77}Si_{10}B_{10}C_3$ microwires have been investigated. The composites containing orthogonal wire arrays display a remarkable transmission window in the frequency band of 1 to 6 GHz under zero external magnetic field indicating an intrinsic double-negative-index characteristic. The polymer matrices have proved to exert a synergistic effect on the microwave properties, which is responsible for the disappearance of the transmission windows when $E_k$ is perpendicular to the glass fiber direction. The plasma frequency of the orthogonal microwire array composite is higher than that of the parallel microwire array with identical wire spacing; this could be attributed to the enhanced microwire-wave interaction induced by the axial electrical components on the additional layer of perpendicular wires. All these features make this new kind of orthogonal microwire composites promising for potential cloaking and sensing applications.

a) Corresponding author: h.x.peng@bristol.ac.uk(HXP)

b) Corresponding author：faxiang.qin@gmail.com(FXQ)

## I. Introduction

Metamaterials have attracted considerable research interest in the past two decades due to their remarkable artificial properties since the pioneering work by Sir Pendry[1] and Smith.[2] So far, studies on conventional metamaterials have focused on seeking for a double negative (DNG) characteristic in a target frequency band by periodically arranging functional units such as the split ring resonator (SRR) and have covered the whole wave spectrum from optical wavelength[3] to microwave wavelength.[4,2] However, the fabrication of conventional metamaterials has been a rather complicated process due to the necessity of constructing and combining the functional units, rendering the metamaterial a 'structure' rather than a true 'material', not to mention the rather limited frequency bands to achieve the DNG characteristics. Recently, ferromagnetic microwires are considered as feasible building blocks of metamaterials due to their excellent soft magnetic properties, giant magnetoimpedance/giant stress-impedance effects and the tunability towards external fields.[5,6] From the multifunctional standpoint, it is natural to realize that by incorporating microwires into high mechanical-performance base materials we would be able to introduce additional electromagnetic properties therein with minimum perturbation to the excellent mechanical properties of the matrix. At this point, it is possible and also of particular interest to use the term of *metacomposite* to consider the above composite system, which by definition is a true material. Unlike conventional metamaterials, the ultimate properties of such metacomposites are dependent not only on the local properties of microwires but also on their geometrical parameters, which offers more flexibility to manipulate the microstructural configuration to meet different engineering requirements.[7]

Fundamentally, it has been confirmed that an array of periodically arranged conductive wires could formulate a left-handed medium in a certain frequency band.[1] It follows that free standing Co-based microwire arrays without any matrix have been proved to exhibit a typical metamaterial behaviour as verified by the transmission window in the microwave band and the flexible tunability towards the external fields.[8,9] Yet Fe-based microwires would be more suitable to construct metacomposites with the metamaterial behavior arising from their natural ferromagnetic resonance (NFMR).[10] They are also more practical than Co-wires from the economical point of view. Mostly recently, we have demonstrated that by incorporating parallel Fe-based ferromagnetic microwire arrays into polymer matrix, peculiar metacomposite characteristics can be obtained.[7] Furthermore, an orthogonal structure consisting of ferromagnetic microwires and conductive fibers was proposed by Liu et al[11] and a double-negative behaviour was theoretically predicted. Hence, we believe that, owing to the excellent magnetic properties of Fe-based microwires, more salient metamaterial features could be attained by employing orthogonal ferromagnetic microwire arrays alone, which also lead to more simplified fabrication process. In this context, we for the first time designed and manufactured glass fiber reinforced polymer composites filled with orthogonal Fe-based microwire arrays. The prepared composites indeed showed remarkable metacomposite characteristics, which will be useful for potential sensing and cloaking applications.

## II. Experimental

In the present experiment, amorphous glass-coated ferromagnetic microwires $Fe_{77}Si_{10}B_{10}C_3$ with a total diameter of 20 μm and glass coat thickness of 3.4 μm were fabricated by Taylor-Ulitovskiy technique and provided by TAMAG, Spain. The continuous microwires were embedded into 950 E-glass epoxy prepregs in an orthogonal manner with a fixed vertical (along the glass fiber direction) spacing of 10 mm and horizontal spacing of 3 and 10 mm, respectively, as schematically shown in Fig. 1(a). Conventional autoclave curing cycle was adapted to yield the resultant orthogonal microwire composites with size of 500×500×1.25 $mm^3$. For comparison, composites containing parallel microwire arrays along glass fiber direction with wire spacing of 3 and 10 mm (Fig. 1(b)) and blank glass fiber composite without Fe-microwires were also fabricated following the same process. The quasi-static magnetic properties were measured using Lake Shore VSM with axis along microwires under magnetic field up to 5k Oe. The tensile properties of microwires were obtained by employing Instron1466 with the load cell of 1 kN and a video gauge capable of precisely capturing the strain.[12,13] The electromagnetic characterization was performed using free space measurement with the electrical component $E_k$ along (Fig. 1) or perpendicular to the glass fiber direction. More experimental details can be found elsewhere.[14] The *S*-parameters were measured in the frequency band of 0.9 to 17 GHz without and with external fields up to 3000 Oe. The complex permittivity was computed through Reflection/Transmission Epsilon Fast Model.

**III. Results and discussions**

The mechanical properties and magnetization curve of the $Fe_{77}Si_{10}B_{10}C_3$ microwire are shown in Fig. 2 together with a SEM image of the microwire. The rectangular magnetic hysteresis (M-H) loop exhibits a coercivity of 0.31 Oe and a saturation magnetization of 850 Gs. The stress-strain curve in the inset of Fig. 2 shows that the tensile strength and fracture strain are 1297 MPa and 2.88 %, respectively. These results indicate that the Fe-based microwires have excellent soft magnetic and mechanical properties, rendering them advantageous as multifunctional sensing and/or reinforcing elements in composite system.

The transmission, reflection and absorption coefficients of composites containing microwire arrays of different configuration measured along the glass fibers in the absence of external field are shown in Fig. 3. Generally, a much larger transmission is displayed for the 10 mm orthogonal microwire composite than the 3 mm orthogonal one because of the lower microwire volume fraction. Clearly, a transmission window can be seen in the frequency band of 1 to 6 GHz for both orthogonal microwire array composites (Fig. 3(a)). As is known, the NFMR frequency of microwires can be calculated via $f_{FMR} = \gamma\,(M_s+H_a/2\pi)$, where $\gamma/2\pi$=2.8MHz/Oe, $M_s$ and $H_a$ (vanishing compared to $M_s$) are gyromagnetic ratio, magnetization saturation and anisotropy field, respectively.[15] We substitute $M_s$=828 Gs as per Fig. 2 into the equation and obtain 2.4 GHz of the $f_{FMR}$. Indeed, from the absorption spectra (Fig. 3(c)), the absorption peaks are identified to be ferromagnetic resonance (FMR) peaks in the low frequency band of 1 to 3 GHz. Hence a negative permeability dispersion can be obtained above the FMR frequency originated from the longitudinal anisotropy field of the used Fe-based microwires (also evidenced in Fig. 2).[5] Meanwhile, the horizontal wire arrays along the $E_k$ component of incident waves (Fig. 1) guarantee the plasmonic behaviour of the composites, rendering a negative permittivity dispersion below the plasma frequency ($f_p$), which is 6.2 GHz for a spacing of 10 mm as evidenced in Fig. 6(a).[1,7] More detailed discussion of the effective permittivity of the microwire composites will be done later. Together, intrinsic magnetic properties of microwires and their alignment in the composite indicate the DNG features and explain the identified transmission window. To our knowledge, in an orthogonal wire composite system, such metacomposite feature has not been reported previously. In contrast, a much lower transmission (with no transmission window) in the corresponding frequency range is obtained for the 10 mm parallel wire array composite but with a reduction of loss. Besides, it is worth mentioning that all the displayed metacomposite features are obtained without the assistance of external field, which provides more degrees of freedom in applying such materials for cloaking devices from an engineering point of view. In another perspective, for the orthogonal wire array with 3 mm spacing, the transmission is bit larger than the 3 mm parallel array composite (Fig. 2(a)). This improvement is logically attributed to the additional wire array perpendicular to $E_k$ (Fig. 1(a)). For such orthogonal configuration, both permittivity and permeability are enhanced. The improved permittivity comes from the dielectric response of perpendicular wires due to some small axial component of the electric field, which also generates circumferential magnetic field that enhances the permeability to a similar extent.[16] However, the extra contribution to permeability from the excitation of magnetic component of microwave field along the additional wire array should be also considered.[17] Thus transmission of the 3 mm orthogonal microwire array composite can be eased through the improved impedance match, taken into account that $Z = (\mu/\varepsilon)^{1/2}$. This enables the possibility of quantitatively controlling transmittance in the orthogonal array composite through the addition of vertical wires.

Apart from FMR peaks, at higher frequencies, additional absorption peaks are detected for both 3 mm orthogonal- and parallel- composites. In the case of parallel wire array composite, the displayed multi-peak feature in 6 to 14 GHz frequency band is likely due to the fact that narrowing down wire spacing to a critical value (between 3 and 10 mm in the present study) can induce the long-range dipolar resonance among the wires.[7] On the other hand, it is indicated in Fig. 3(c) that the addition of 10 mm spaced vertical wires to form the orthogonal architecture can dramatically reduce the critical wire spacing to below 10 mm and induce microwave interaction peaks at 7.5 GHz. Also, a similar transmission dispersion between the orthogonal microwire composite and the blank composite (without microwires) is noted (Fig. 3(a)), which implies that the matrix material has a synergistic effect on the microwave behaviour of composite.

To clarify this issue, we present in Fig. 4 the transmission and absorption dispersion after mathematically subtracting the S-parameters of the blank composites to consider the influence of matrix material. Generally, the matrix has significant impact on absorption features of microwire composite. Noting that here the $E_k$ was placed along the glass fiber direction, we believe that, as there is not much direct contact of horizontal microwire array and the glass fiber is non-conductive, the interface between wires and glass fibers themselves can be regarded to have zero contribution to the electromagnetic responses. Therefore, in principle, the effect of matrix material on the microwave performance of the microwire composite comes from two aspects: the polymer in the matrix and the interface between microwires and polymers. It should be addressed that the polymer matrix itself does not evidently affect the microwave characteristics of the composites due to their rather limited dielectric permittivity (1~ 3).[18] This is also evident from the absorption spectrum of the blank composite (Fig. 3(c)). In addition, our previous work[19] has demonstrated that the local properties of microwires such as domain structure can profoundly influence the composite absorption features. Owing to the temperature and stress sensitivity,[13,15,20] the microwire domain structure could have been modified via the microwire-polymer interface condition change during the curing process through the applied heat and the transferred mechanical stress from the polymer, finally leading to the absorption variation (Fig. 4(b)). The results also show that the transmittance of a composite is slightly larger than the single orthogonal microwire array structure without a matrix phase (Fig. 4(a)), revealing a positive influence of matrix on metacomposite features. In order to achieve a given transmittance level, less wire amount would be needed for a wire composite than a pure wire array. This suggests that the metacomposite features are influenced not only by intrinsic properties and topological arrangement of microwires, but also by the existence of the polymer matrix and the related interface therein. Note that the above synergistic effect is revealed providing that $E_k$ is along the glass fiber direction.

To further recognize the effect of glass fibers on microwave behaviour from the matrix, we present in Fig. 5(a) the transmission coefficients of the 10 mm orthogonal microwire array composite when $E_k$ is along and perpendicular to glass fibers, respectively. Remarkably, the transmission window disappears for the latter case. Considering the glass fibers are unidirectional, arranging microwires along the glass fibers can contribute to the DNG features through the preservation of the continuous microwire configuration. In this scenario, microwires are more likely to assign themselves into the gaps among in-prepreg and/or between-layer glass fibers, making the classic dilute plasma medium theory and assumptions on microwires of perfect ferromagnetic components still valid. On the contrary, the hard physical contact between glass fibers and vertical microwires would lead to considerable stress concentration on the wire surface (Fig. 5(b)), even the bending of the microwire after curing. When measuring perpendicularly to glass fibers, this stress can not only impact magnetic response arising from the modification of domain structure, making the NFMR characteristic hard to predict, but also possibly damage the plasmonic behaviour of continuous vertical microwires, rendering the negative permittivity impossible. Therefore, wire-orientation can profoundly impact on the observed DNG features, i.e., metacomposite characteristics can be turned on/off by simply rotating composite by 90 degrees. Note that the above on/off behaviour can be further extended to an angle-dependent metamaterial behaviour, giving the possibility of mechanically tuning the fabricated metacomposites. This is reminiscent of the recent studies either by restricting liquid crystal in the certain periodic structure,[21,22] or employing resistor network based on transformation optics theory to electrically tune the metamaterials.[23] However, disadvantages of complexity in manufacturing and the burden of extra DC power source in these studies cannot be ignored. From above reasoning, we have displayed an orthogonal microwire metacomposite capable of being tuned effectively via mechanically rotating samples. This indicates a possible new path for the dynamic control of electromagnetic wave's propagation in the metamaterial devices. Future work is necessary to quantify relationship between the metamaterial behaviour and the microwire-orientation angles.

So far we have addressed the DNG characteristics and the influences from the matrix material. To look further into the underlying physics of the studied composites, we present the complex permittivity of experimentally observed orthogonal array microwire composite in Fig. 6. It is well established that, for conductive wire array, the plasma frequency ($f_p$), can be determined as $f_p^2 = \frac{c^2}{2\pi b^2 ln\left(\frac{b}{a}\right)}$, where $a$ and $b$ are wire diameter and spacing, respectively.[1] Using above equation, one can receive the $f_p$ of 4.8 GHz and 16.6 GHz for the parallel wire array composite with spacing of 10 mm and 3 mm, respectively. However, as indicated in the figure, the plasma frequencies are increased respectively to 6.1 GHz and over 17 GHz for the two composites. This is also consistent with the observed transmission windows associated with DNG features in the corresponding frequency band. As discussed before, the overall dieletric response of the orthogonal composites can be enhanced via the microwave interactions between vertical wires and small axial component of $E_k$. In other sense, we can consider the vertical wires as the short-cut dielectric components among adjacent horizontal microwire and the small amount of the increased $f_p$ can be attributed to the extra response from these equivalent short wires. It is worth remarking that, the plasma frequency can be further increased by adding more vertical wires into the composite would elevate the $f_p$ or performing a treatment such as stress/current annealing on the wires prior to incorporation into polymer matrices.[5,24]

**IV. Conclusion**

In summary, we have designed and fabricated an orthogonal microwire metacomposite, which shows remarkable natural double negative features. The composite architecture including the wire-spacing and the matrix material prove to be responsible in the synergistic and anisotropic electromagnetic manner for the observed metacomposite behaviour in terms of operation frequency and transmission level. The particular merit of the orthogonal structure, as compared with the parallel composite with the same wire-spacing, lies in the fact that the perpendicular wires can benefit the transmission level with better impedance match and blueshift plasma frequency with the enhanced dynamic microwire-wave interaction.

**Acknowledgements**

Yang Luo would like to acknowledge the financial support from University of Bristol Postgraduate Scholarship and China Scholarship Council. FXQ was supported under the JSPS fellowship and Grants-in-Aid for Scientific Research No. 25-03205.

**Figure captions:**

Fig. 1 (Colour online) Schematic view of manufacturing process of (a) orthogonal wire array composite with fixed wire spacing 10 mm perpendicular to glass fiber direction and different horizontal wire spacing of 3 and 10 mm, respectively and (b) parallel wire array composite with wire spacing of 3 and 10 mm, respectively.

Fig. 2 (Colour online) Quasi-static magnetization curve of $Fe_{77}Si_{10}B_{10}C_3$ microwires. The insets are cross-sectional SEM image and stress-strain curve of the microwire, respectively.

Fig. 3 (Colour online) Frequency plots of (a) transmission, (b) reflection and (c) absorption spectra of composites with parallel and orthogonal wire arrays with $E_k$ along glass fibers in the absence of external magnetic fields.

Fig. 4 (Colour online) Frequency plots of (a) transmission and (b) absorption spectra of orthogonal microwire array and their composites after eliminating the influence of glass fibers.

Fig. 5 (Colour online) (a) Frequency plots of transmission dispersion of orthogonal microwire array composite when $E_k$ is perpendicular to glass fibers and (b) schematic view of vertical and horizontal microwires in the polymer matrices.

Fig. 6 (Colour online) Frequency plots of (a) real part $\varepsilon'$ and (b) imaginary part $\varepsilon''$ of permittivity of the orthogonal microwire array composites. Inset in (a) is a zoomed view of part of the curve inside the square mark.

(a)

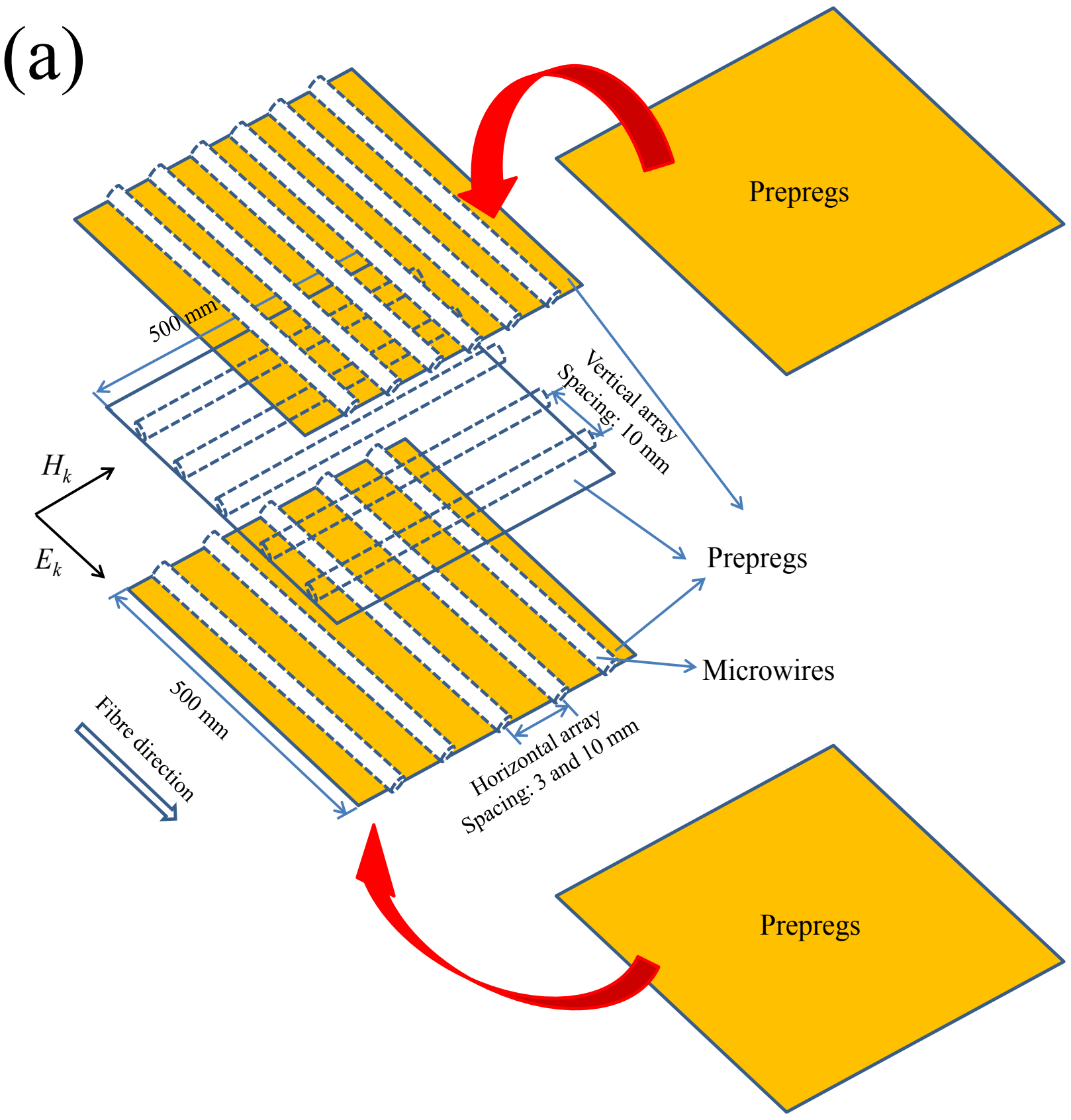

(b)

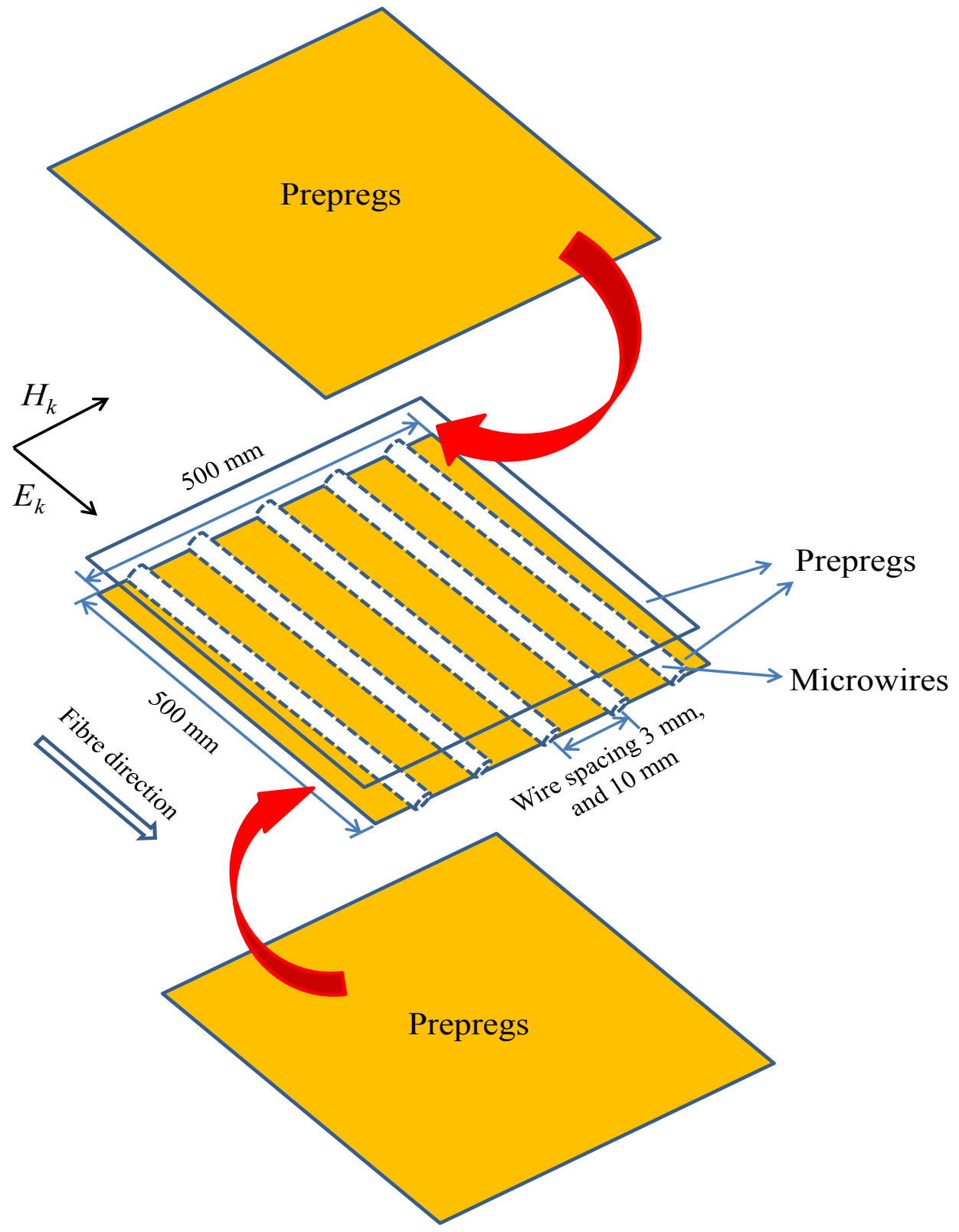

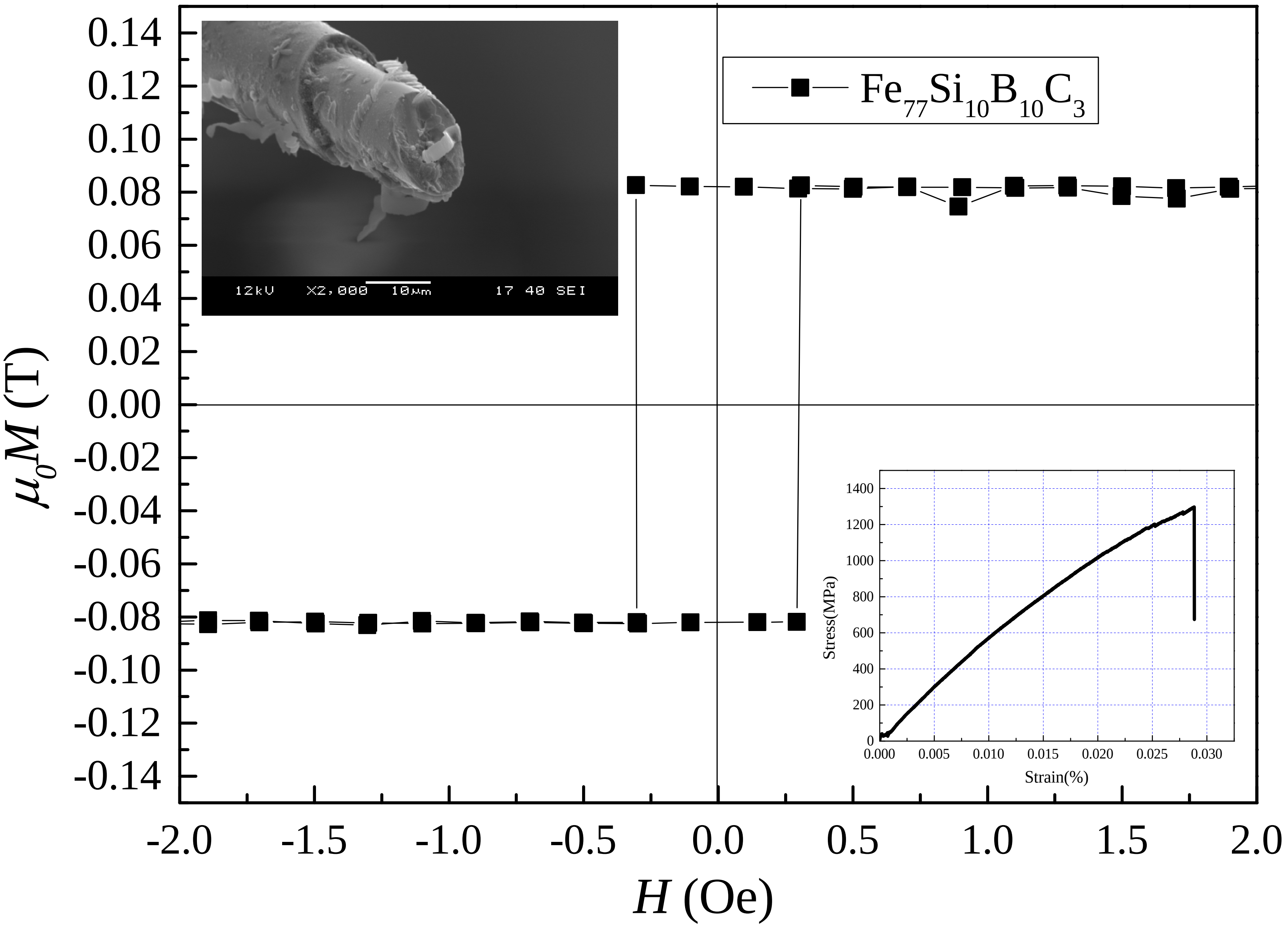

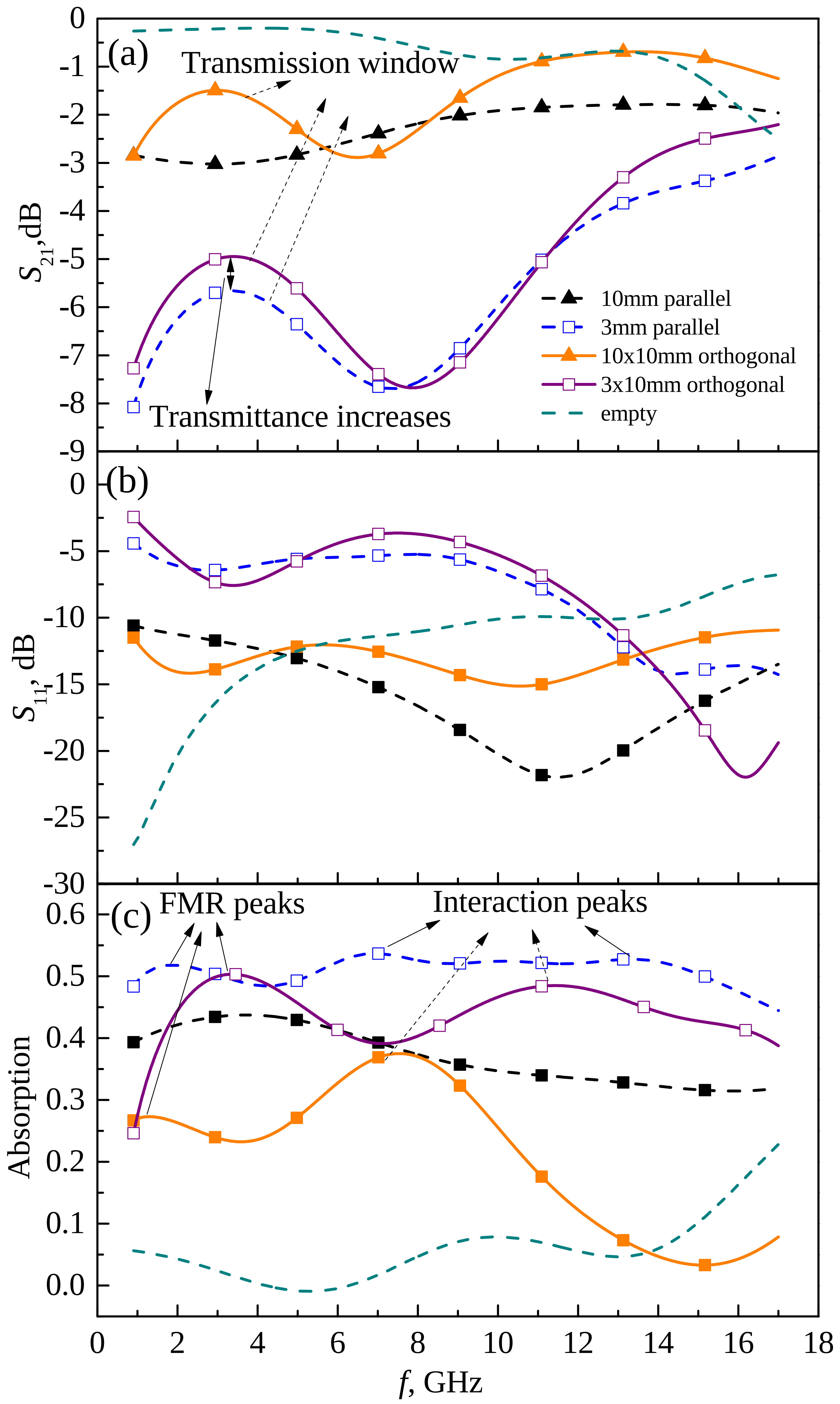

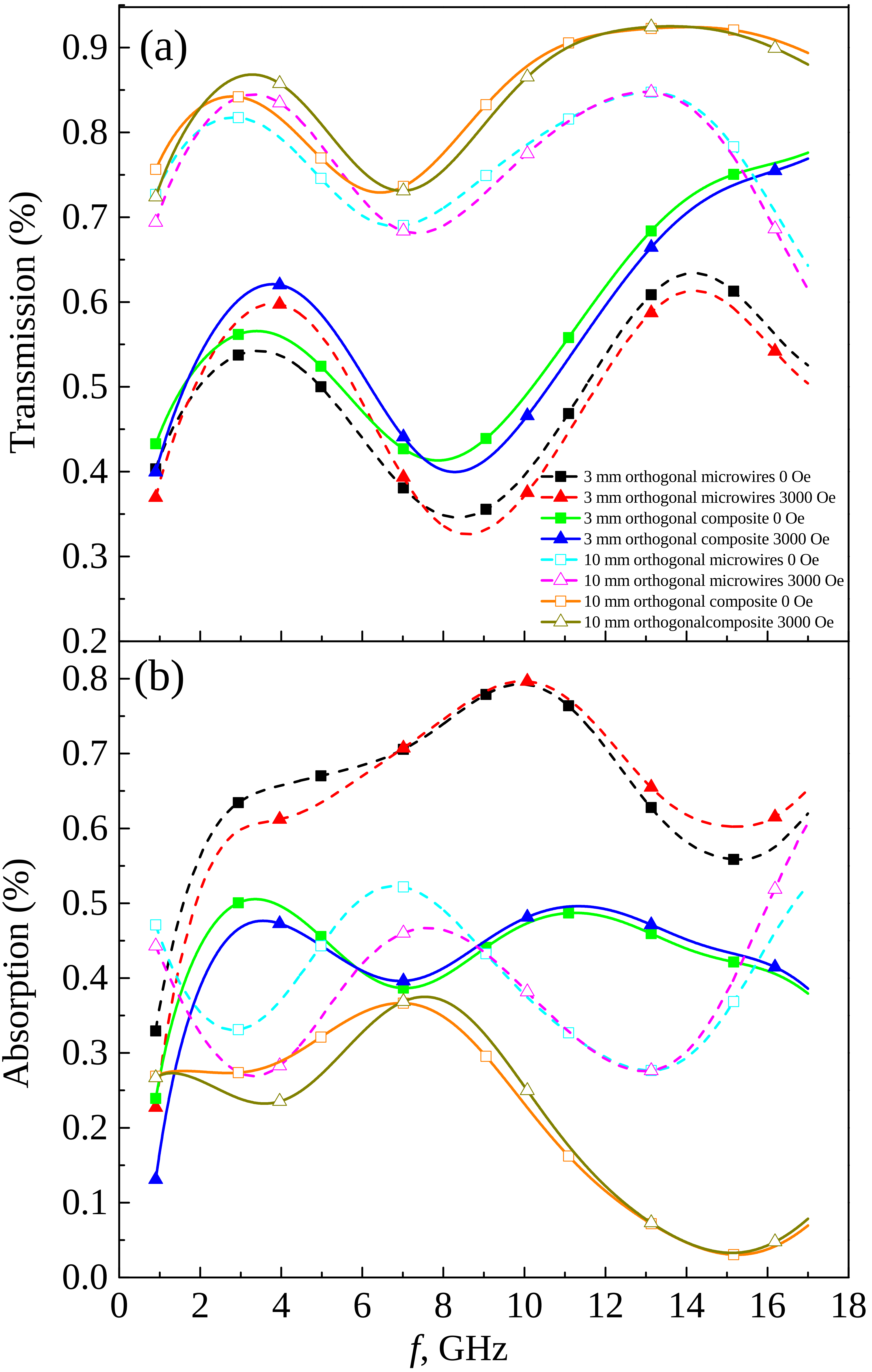

(a)
(b)
Transmission (%)
Absorption (%)
f, GHz
0.9
0.8
0.7
0.6
0.5
0.4
0.3
0.2
0.1
0.0
0
2
4
6
8
10
12
14
16
18
3 mm orthogonal microwires 0 Oe
3 mm orthogonal microwires 3000 Oe
3 mm orthogonal composite 0 Oe
3 mm orthogonal composite 3000 Oe
10 mm orthogonal microwires 0 Oe
10 mm orthogonal microwires 3000 Oe
10 mm orthogonal composite 0 Oe
10 mm orthogonalcomposite 3000 Oe

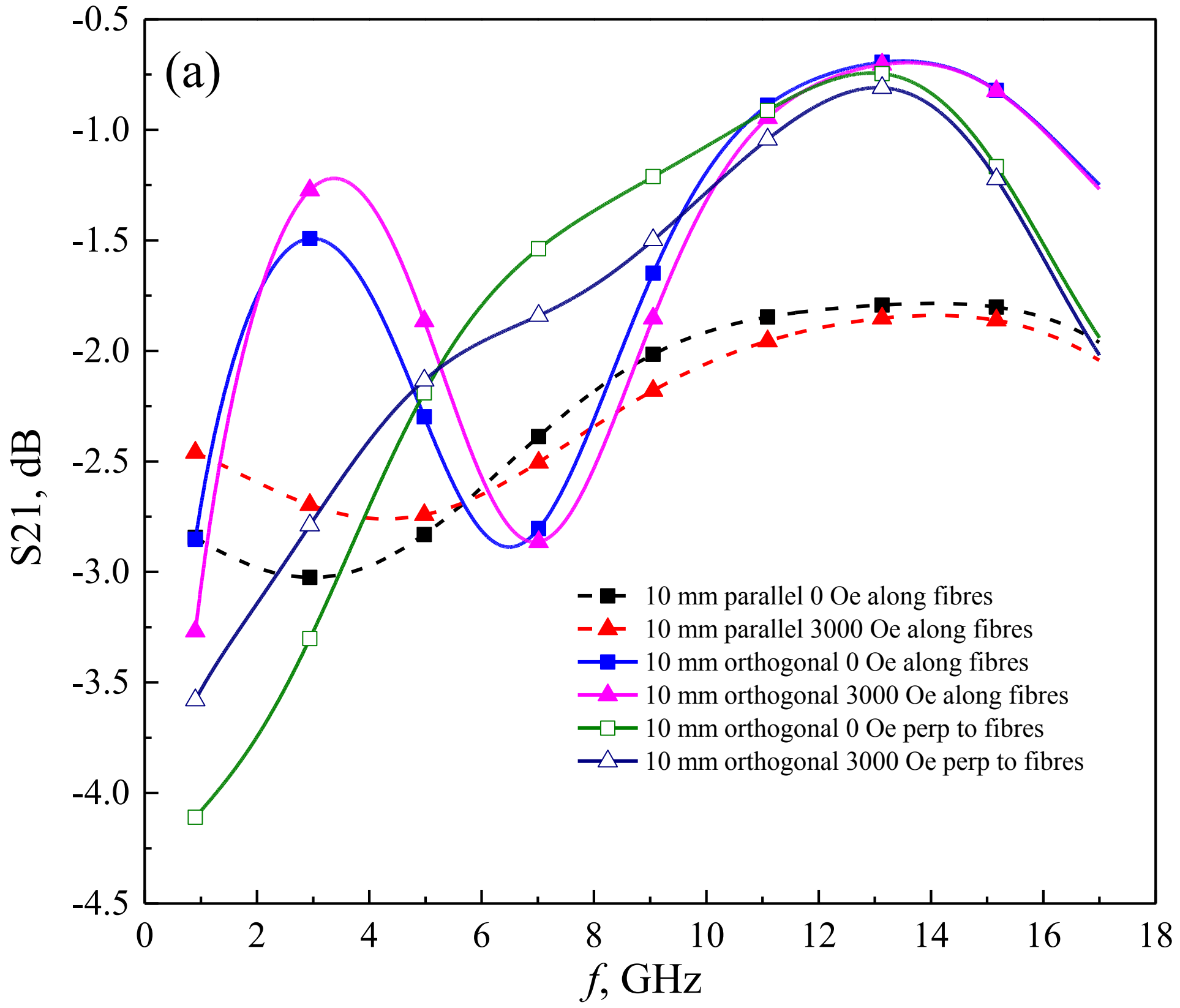

(a)
S21, dB
f, GHz
-0.5
-1.0
-1.5
-2.0
-2.5
-3.0
-3.5
-4.0
-4.5
0
2
4
6
8
10
12
14
16
18
10 mm parallel 0 Oe along fibres
10 mm parallel 3000 Oe along fibres
10 mm orthogonal 0 Oe along fibres
10 mm orthogonal 3000 Oe along fibres
10 mm orthogonal 0 Oe perp to fibres
10 mm orthogonal 3000 Oe perp to fibres

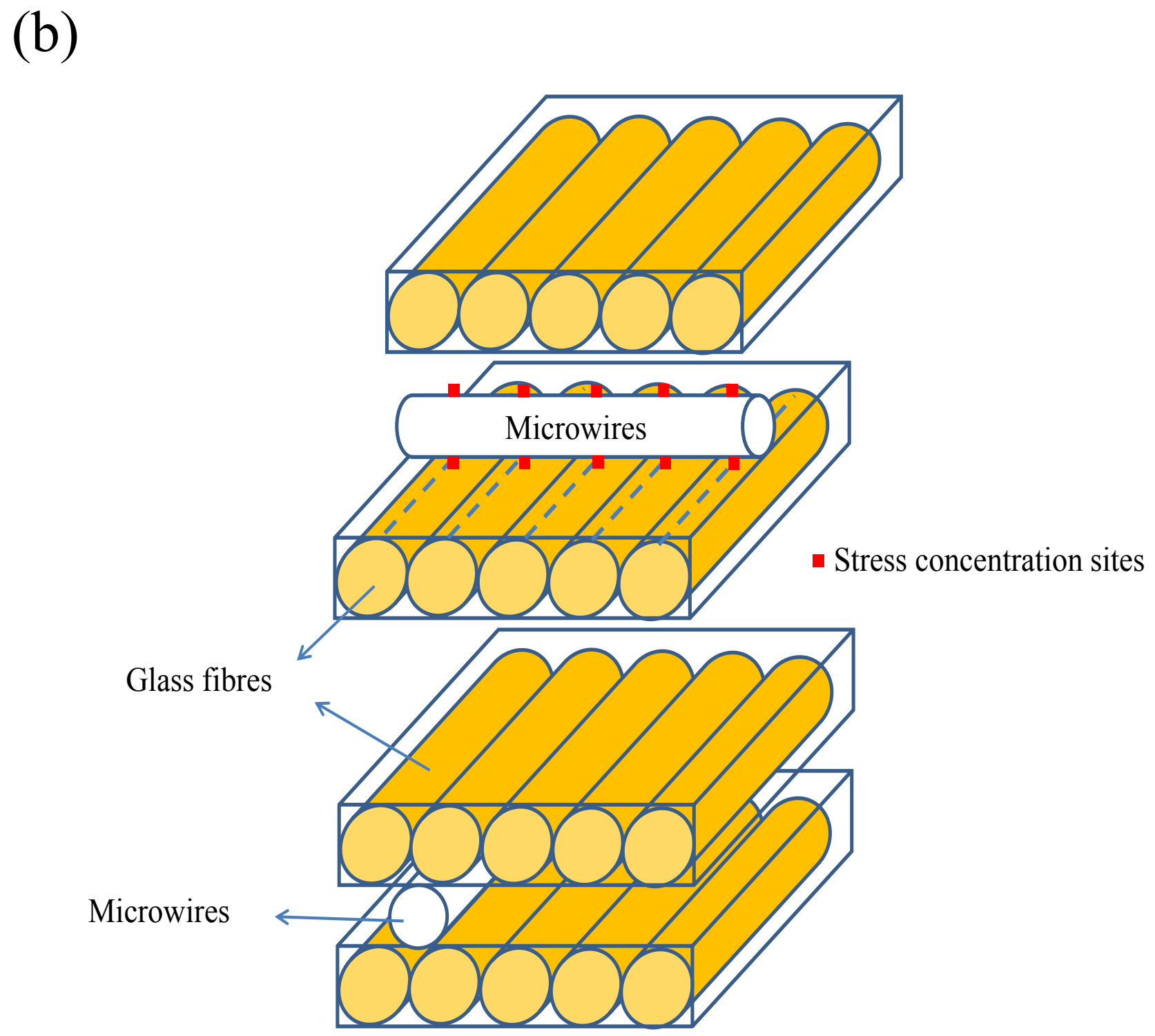

(b)
Microwires
Stress concentration sites
Glass fibres
Microwires

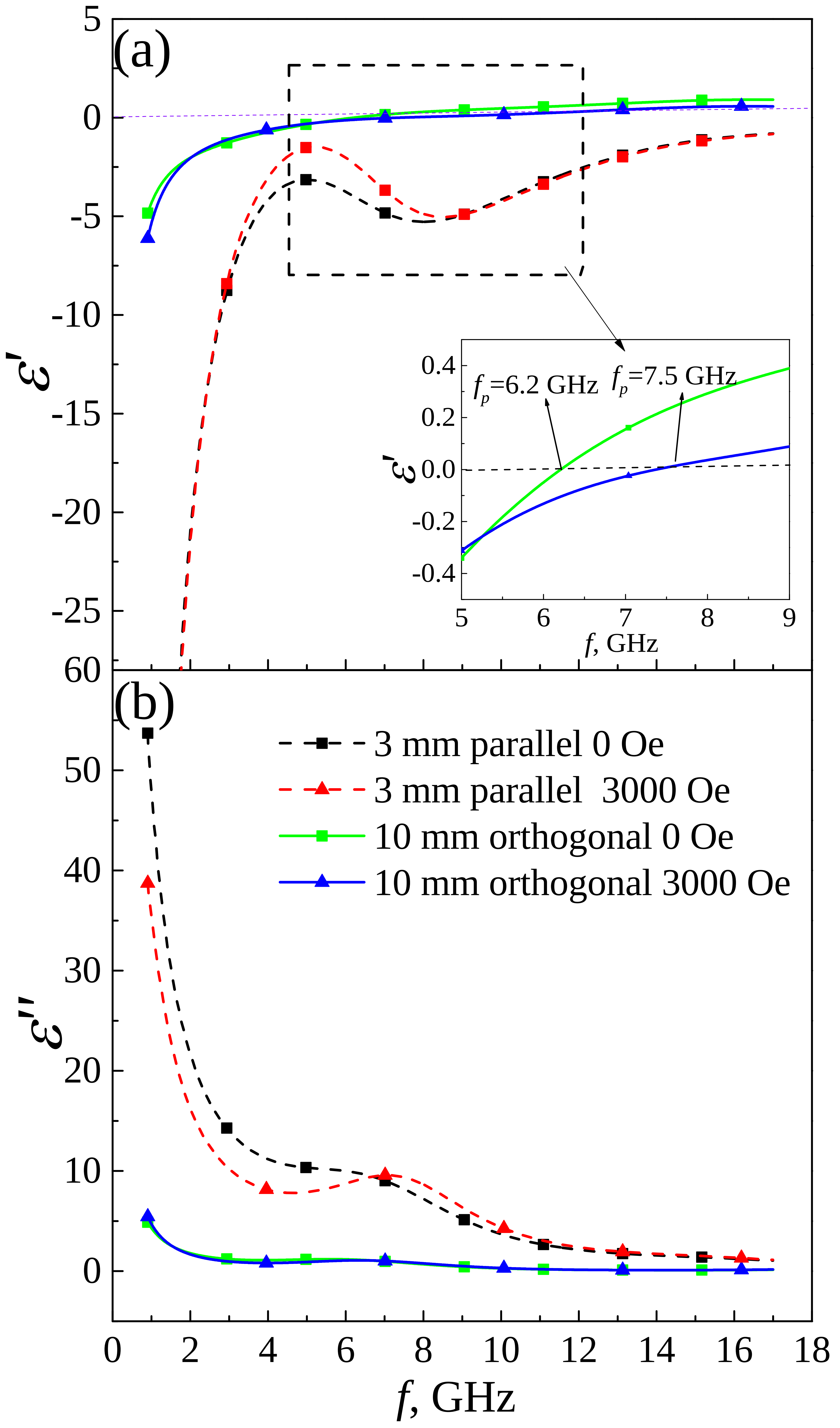

(a)
(b)
ε'
ε''
f, GHz
$f_p$=6.2 GHz
$f_p$=7.5 GHz
3 mm parallel 0 Oe
3 mm parallel 3000 Oe
10 mm orthogonal 0 Oe
10 mm orthogonal 3000 Oe